# Impact of excess and disordred Sn sites on $Cu_2ZnSnS_4$ absorber material and device performance: A $^{119}$Sn Mossbauer Study

Goutam Kumar Gupta[1], V R Reddy[2],and Ambesh Dixit[1,a]

[1]Department of Physics &Center for Solar Energy, Indian institute of technology Jodhpur, Rajasthan, 342037 India
[2]UGC-DAE CSR Indore Center, Indore, 452017 India

[a)]Corresponding author: ambesh@iitj.ac.in

**Abstract**

Mossbauer analysis is carried out on CZTS samples, subjected to a low temperature processing at $300^0$C (S1) and high temperature processing at $550^0$C under sulfur environment (S2). Loss of Sn is observed in sample S2 due to high temperature thermal treatment.The isomer shifts obtained in the Mossbauer spectra confirms the existence of Sn at its 4+ valance state in both the samples. Relatively high quadriple splitting is observed in S1 with respect to S2, suggesting dislocations and crystal distortion present in S1, which are reduced drastically by high temperature annealed S2 sample. The fabricated solar cell with S1 and S2 absorbers showed significant improvement in efficiency from ~0.145% to ~1%. The presence of excess Sn in S1 allows enhanced recombination and the diode ideality factor shows larger value of 4.23 compared to 2.17 in case of S2. The experiments also validate the fact that S1 with Sn rich configuration shows lower acceptor carrier concentration as compared to S2 because of enhanced compensating defects in S1.

**Keywords:** $^{119}$Sn Mossbauer spectroscopy, CZTS compound semiconductor, Solar Cell; Valence state, Defects

**Introduction**

Kesterite material $Cu_2ZnSnS_4$, also named as CZTS, has emerged as a promising photovoltaic (PV) absorber layer material. Its attractive optical and electronic properties such as high absorption coefficient and optimum direct band gap make it a suitable alternative for silicon photovoltaics. In addition, this is considered as a good replacement for existing chalcopyrite $CuInGaSe_2$ (CIGS) absorber, which aready demonstrated a large photovoltaic efficiency of about 22.6%[1]. Kesterites bear the advantages of their low cost earth abundant constituents, unlike chalcopyrite which suffers with indium scarcity and gallium (Ga) toxicity. Further, the bandgap tunability by alloying with selenium favours the optimization of bandgap to match the solar spectrum, realizing the enhanced photoconversion efficiency. Several vacuum and non-vacuum based deposition methods are optimized till date to fabricate CZTSSe based solar photovoltaic devices with significant efficiencies[2][3][4][5][6]. However, the obtained maximum efficiency is still far behind Schockley-Quieser limit (~32.8%) for CZTS,Se solar cell[7]. The pure sulfide CZTS and selenide CZTSe based photovoltaic devices reached the maximum efficiency upto 9.2% and 11.6% with co-evaporation techniques[8][9]. Further, sulfoselenide CZTSSe absorber based photovoltaic cell showed the maximum efficiency ~12.6% and ~12.7% with single CdS buffer layer and with double $CdS/In_2S_3$ buffer layer using hydrazine based pure solution approach[10][11]. Kesterite solar devices commonly suffer from the problem of volage deficit which presents significant challenge in realizing the high efficiency[12][13]. Recently Priscilla D. Antunez et al. reported near maximum efficiency with high open circuit voltage, using exfoliation of back contact from the substrate, used for growth and then depositing the high work function back contact on the grown device [14]. Incorporation of Ge in CZTS,Se also shows significant possibility to get rid of this voltage deficit problem [15]. There are numerous materials and device issues and challenges to realize high photovoltaic efficiencies with CZTS. For example being a quaternary material, this system is prone to acceptor and donor type defects and has a very narrow region of phase stability. Defects in the CZTS system includes vacancies ($V_{Cu}$, $V_{Zn}$, $V_{Sn}$, $V_S$), antisites ($Cu_{Zn}$, $Cu_{Sn}$, $Zn_{Sn}$, $Zn_{Cu}$, $Sn_{Cu}$, $Sn_{Zn}$) and interstitials ($Cu_i$, $Zn_i$, $Sn_i$, $S_i$)[16]. Relatively lower formation energy of acceptor defects compared to that of donor defects, makes CZTS a p type material without any foreign element doping. Stochiometric CZTS thin film shows inferior performance compared to the non-stoichiometric films in terms of photovoltaic response[17]. However non-stoichiometry sometimes may lead to the high density of defects and secondry phases, simultaneously, causing degraded performance[18][19].

Zn and Cu disorders are quite common because of their similar chemical structures, such as ionic radii. These two atoms can easily swap at lower enthalpic cost to form $Cu_{Zn}$ acceptor and $Zn_{Cu}$ donor antisite defects[20][21]. Vacancy of Cu also introduces acceptor level and $V_{Cu}$ and $Cu_{Zn}$ are dominant acceptor defects in CZTS system. These defects in conjunction $[Cu_{Zn} + Zn_{Cu}]$ and $[V_{Cu} + Zn_{Cu}]$ form defect complexes, which are self compensating and with comparatively lower formation energy of $Cu_{Zn}$ and $V_{Cu}$ acceptor defects[22]. These defects led to p-type CZTS material. $Cu_{Zn}$ has acceptor level deeper inside the band gap compared to $V_{Cu}$. This suggests that acceptor carrier concentration in CZTS due to $Cu_{Zn}$ is not optimal. It is also observed that even in stoichiometric CZTS, $Cu_{Zn} + Sn_{Zn}$ and $2Cu_{Zn} + Sn_{Zn}$ defect clusters exhibit high concentration causing significant bandgap decrease and are detrimental to the device performance[23]. Therefore, Cu deficient and Zn rich configuration are suggested to be favorable for the high performing devices[24][21]. However, any non-stochiometry in CZTS allows formation of unwanted binary and ternary impurity phases. Cu and Sn rich CZTS favors formation of $Cu_2S$, SnS, and $Cu_2SnS_3$ binary and ternary impurity compounds. These impurities are conductive and provide shunting path in the device, thus decreasing the open circuit voltage of the device. Zn rich CZTS favors formation of ZnS impurity which is detrimental but passivate the probable shunt path present in the CZTS. Sn in the CZTS lattice exits in +IV sate at its native site. However, Sn in CZTS can replace Cu and Zn from their lattice sites and can create antisite point defects $Sn_{Cu}$ and $Sn_{Zn}$, deep in the bandgap and their high formation energy make them relatively uninfluential[23]. Further, multivalent Sn can exist in +II or +IV oxidation states. Sn at the Cu site $Sn_{Cu}$ exists exclusively as a divalent atom and forms shallow single donor level[25] and Sn at the Zn site either forms double donor or isoelectronic center related to its multivalent nature. Isoelectronic centers, formed due to change in valancy of Sn from IV to II state, acts as non radiative recombination centers, while at its +IV valance state it forms deep donor level far from the conduction band and hence serves as the recombination center and not as shallow donor. These defects in CZTS severly affect the photovoltaic response and hence should be avoided by taking care of nonstochiometry in CZTS[25].

In this paper, we report the effect of post annealing treatment on structural optical and the electronic properties of sol-gel derived CZTS thin films. The Mossbauer spectroscopic measurements are carried out to understand the local states of Sn ions in CZTS system and their impact on photovoltaic response. Temperature dependent resistance measurements are used to understand the activation energy of defects in CZTS samples. The prepared CZTS thin films are further integrated in Al:ZnO:iZnO/CdS/CZTS/Mo/Glass solar photovoltaic device structures, where Al:ZnO is aluminum doped zinc oxide and served as transparent conducting oxide top electrode for electron collection, i-ZnO is intrinsic zinc oxide, served as window layer, CdS is n-type buffer layer on CZTS absorber layer and Mo is molybdenum, used as back electrode. The fabricated devices are characterized for electrical and impedance properties to understand the photovoltaic response against process parameters.

**Experimental details:**

CZTS thin films are prepared using non-vacuum sol-gel derived spin coating process on 1″ x 1″ soda lime glass (SLG) substrates. These SLG substrates are ultrasonically cleaned in trichloro ethylene (TCE), followed by cleaning in acetone and methanol solutions and finally dried with nitrogen before any thin film deposition. Sol-gel derived spin coating process is used for non-vacuum CZTS thin film deposition. The solution for the spin coating is prepared in 2- methoxy ethanol solvent using chloride salts of constituent metals and thiourea as the sulfur source. Elemental precursors are taken in the ratio of 2:1.2:1:8 of Cu:Zn:Sn:S. Excess Zn in the prepared solution is taken to maintain the Cu poor and Zn rich configuration required for the high performing CZTS devices. Thiourea is also taken in excess to dissolve the metal salts completely and to achieve a highly stable solution. The synthesis details are discussed in detail in one of our previous works[26]. The highly stable sol is then used for preapring CZTS thin films using a spin coater. The spinning is carried out at around 3500 rpm followed by baking at 300 $^0$C on hot plate in open ambient for 5 mins. This spinning and drying process is repeated several time to get the desired thickness. The CZTS sample thus prepared is later kept at $300^0$C in open atmosphere for 30 min and named as S1 for later discussion. An identical sample is annealed at 550 °C under 5% $H_2S$+Ar dynamic gaseous environment in a tubular single zone furnace to avoid any sulfur loss from the system. The heating profile is reported earlier [27]. This sample is named as sample S2 for future discussion.

**Characterization**

The crystallographic structure of these CZTS thin films is analyzed using a Bruker D8 advance X-ray diffractometer (XRD) from 20° to 80° 2θ range. The copper $K_\alpha$ with wavelength 0.15406 nm is used as incident radiation in thin film geometrical configuration. An EVO 18, especial edition, Carl Zeiss scanning electron microscopy (SEM) is used for microstructural measurements and energy dispersive X-ray analysis (EDX) measurements are carried out for the elemental composition analysis using OXFORD instrument EDX accessory, attached with SEM instruments. In addition, atomic force microscopic (AFM) measurements are carried out to understand the surface morphology as a function of growth and post growth treatment for these CZTS samples. Absorbance spectra are recorded to understand the electronic properties using carry 4000 UV-Vis spectrophotometer from 200 nm to 900 nm wavelength range. The room temperature $^{119}$Sn Mossbauer spectroscopic measurements are carried out in conversion electron mode using homemade gas flow counter and operating the spectrometer in constant acceleration mode. The velocity scale is calibrated with respect to natural iron using $^{57}$Fe Mossbauer measurements. The reported Sn- hyperfine parameters are referenced with respect to the powder $SnO_2$ absorber.

**Result and Discussion**

The measured XRD diffractographs are shown in Fig. 1 and the observed diffracted lines are indexed with kesterite CZTS phase (ICDD # 26-0575). The films are textured along (112) planar orientations and the crystallinitiy improves for high temperature treated sample S2. The improved intensity and reduced full width at half maxima (FWHM) of diffraction peaks for S2 CZTS sample, suggest enhancement in the crystallinity with increased grain size. These measurements imply that annealing in $H_2S$ environment at elevated temperature provides enough thermal energy to crystallize CZTS into larger grains. The average crystallite size (D) is calculated using Scherrer formula $D = K\lambda/\beta \cos\theta$ where $K = 0.94$, $\lambda$ is wavelength of characteristic Cu $K\alpha$ X-ray incident radiation, $\beta$ is full width half maxima (FWHM) of the diffraction peaks in radian, and $\theta$ is respective diffraction position. The estimated average crystallite sizes are 77 Å, and 143 Å for samples S1, and S2, respectively. This controlled thermal treatment also assisted in reducing the dislocation density, which is calculated from the estimated average crystallite size by applying Williamson and Smallman equation $\delta = 1/D^2$, where D is the average crystallite size[28]. The calculated dislocation density are $1.68 \times 10^{16}$ lines/m$^2$ and $4.89 \times 10^{15}$ lines/m$^2$ for sample S1 and S2, respectively.

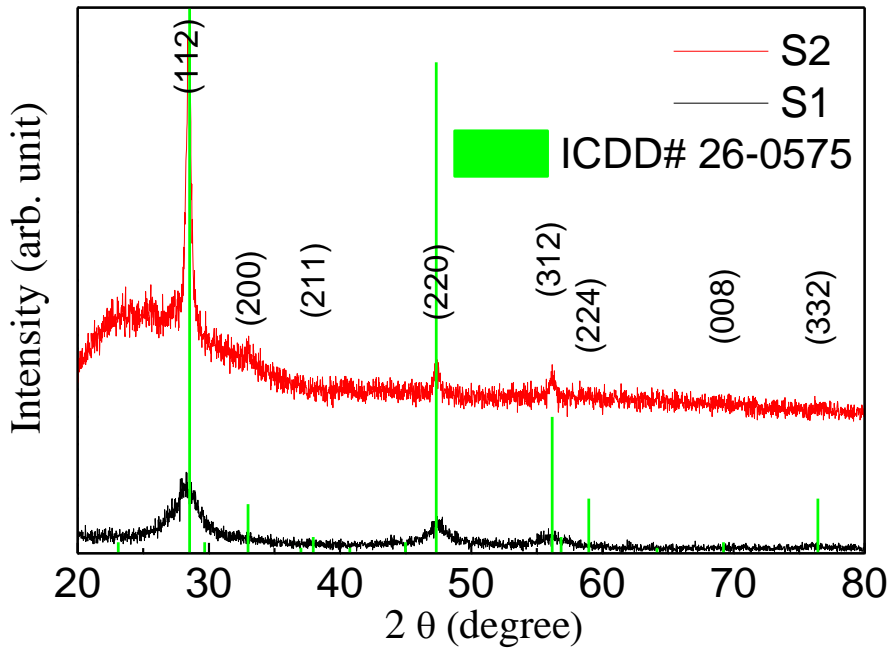

Figure 1: XRD patterns for sample S1, S2 (Green color lines indicate the peak positions of CZTS according to ICDD # 26-0575).

The microscopic and topographic measurements are summarized in Fig 2 and suggest that films are smooth, dense and have large grains. The solution processed S1 and S2 samples are nearly free from any cracks and voids, Fig 2(a & b). EDAX results are summarized in Table 1 for different elements in different CZTS samples. The elemental composition of these samples suggest copper poor and slightly tin rich configuration for S1 sample. Annealing at elevated temperature resulted in the loss of tin from the samples and is relatively poor in tin composition. The stoichiometry changes after post annealing treatment is attributed to the tin loss and sample S2 shows copper poor and zinc rich configuration. Both sample S1 and S2 are showing slighty excess sulfur content from stoichiometric ratios, however the numbers are within the error limits of EDX measurements.

Table 1. Elemental composition for S1 and S2 samples, as extracted from EDX measurements, in conjunction with different atomic ratios

| Samples | Cu | Zn | Sn | S | Cu/Zn+Sn | Zn/Sn | Cu/Sn | Cu/Zn |
|---------|-------|-------|-------|-------|----------|-------|-------|-------|
| S1 | 21.72 | 12.51 | 13.41 | 52.36 | 0.838 | 0.93 | 1.62 | 1.74 |
| S2 | 22.06 | 12.56 | 11.87 | 53.50 | 0.903 | 1.06 | 1.86 | 1.76 |

The controlled copper deficiency is important to realize the p-type electrical conductivity of CZTS material, an important parameter for designing high efficiency solar photovoltaic devices. The surface morphology shown in Fig 2 (a), of S1 suggest dense but smaller grains which after annealing increases in S2 Fig 2(c). This is also in agreement with the observed improved crystallinity S2 compared to S1 in the XRD micrograph. Topographic images of the prepared samples S1 and S2 are obtained using atomic force microscopy (AFM) and is shown in Fig 2 (b, d). The roughness of the samples remain relatively immune with the annealing treatment. The observed RMS roughness , average roughness and grain length summarized in Table 2.

Table 2. Average, RMS roughness and grain size for S1 and S2 samples

| Samples | Rq (nm) | Ra (nm) | Grain (nm) |
|---------|---------|---------|------------|
| S1 | 53.138 | 41.96 | 244.3 |
| S2 | 50.1 | 40.5 | 344 |

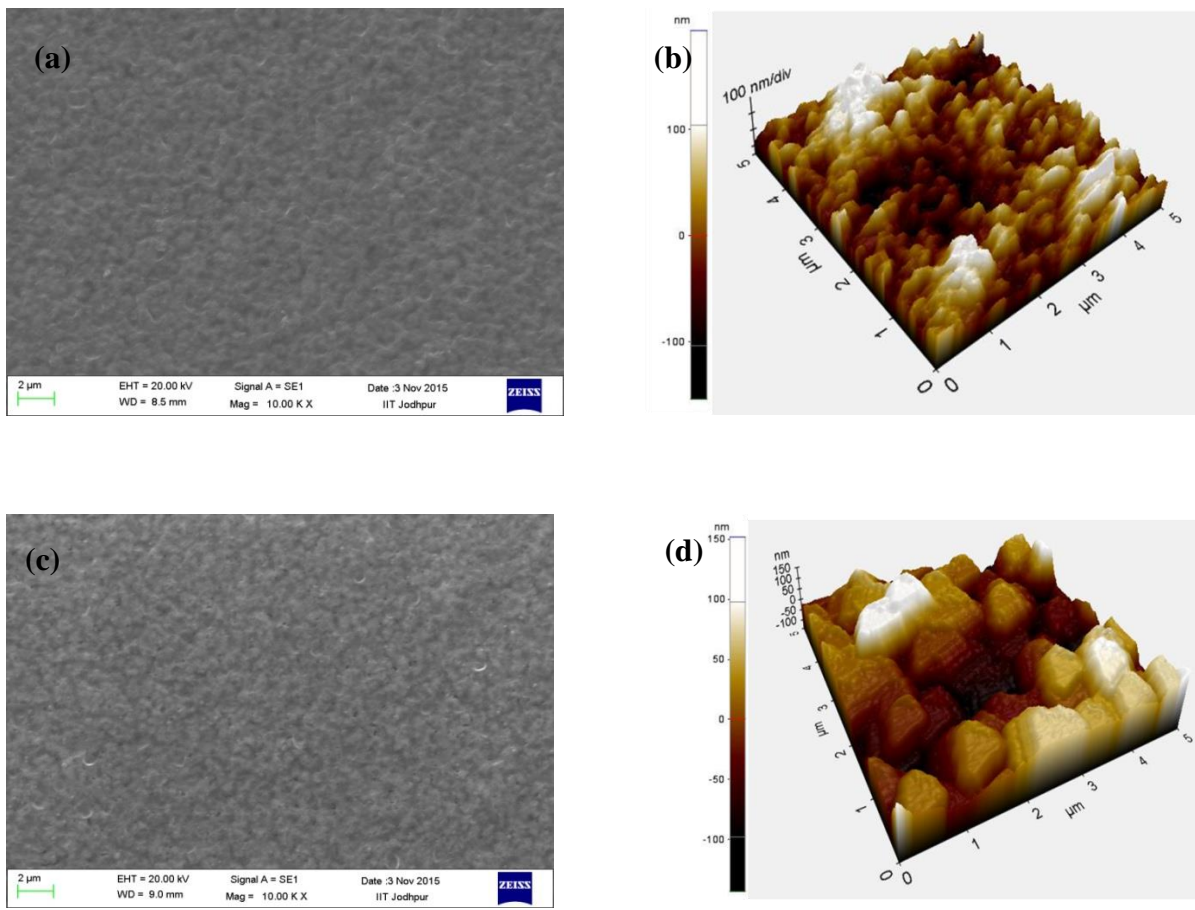

Figure 2: SEM micrograph for (a) S1 and (b) S2 surfaces; and respective topographic images (c) S1 and (d) S2 samples

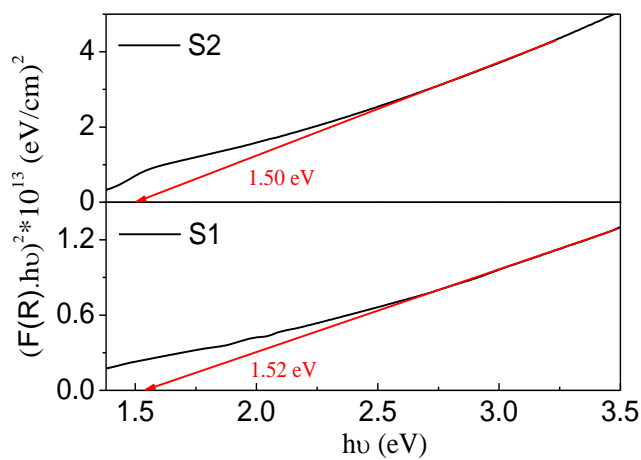

Figure 3: Tauc plot for CZTS thin film samples S1 and S2; used for estimating the band gap values

The optical properties are investigated using UV-Vis reflectance data for these samples. The measured reflectance is used to calculate the spectral absobance F(R) using Kubelka-Munk model, $F(R) = (100-R)^2/2R$ and finally $(F(R).h\upsilon)^2$ as function of energy ($h\upsilon$) is plotted in Fig 3 for S1 and S2 samples. The absorption edge is evaluated from the extrapolated linear region of Tauc plot to evaluate the optical band gap using $(F(R)h\upsilon)^2 = A(h\upsilon - E_g)$ relation, where $Eg$ is the optical band gap of the material. The measured band gap values are close to 1.5eV for both CZTS samples (S1 and S2), as explained using red arrow lines in Fig. 3.

**Mossbauer spectroscopic measurements**

$^{119}$Sn Mossbauer spectroscopy measurements are carried out in transmission geometry for both pristine S1 and post annealed S2 CZTS thin film and the results are summarized in Fig 4.

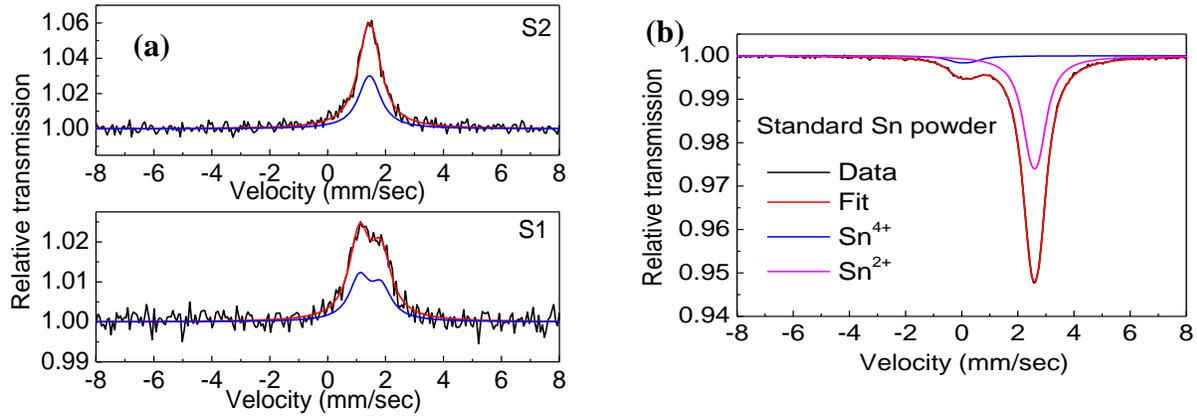

Figure 4: $^{119}$Sn Mossbauer spectroscopic measurements on S1 and S2(a) CZTS samples, with standard Sn powder (b) as a reference.

The data is fitted with a doublet for both S1 and S2 CZTS samples. The measured hyperfine parameters are shown in Table 3. In stoichiometric CZTS, oxidation states of Cu, Zn, Sn and S are +I, +II, +IV, and –II, respectively. However, due to multivalancy of Sn it can exist in both +II and +IV oxidation states. This can be resolved unambiguously from the measured isomer and quadrupole shift values.

Table 3: Isomer and Quadrupole shifts with tin valance state, derived from $^{119}$Sn Mossbauer data

| Sample | IS (mms$^{-1}$) | QS(mms$^{-1}$) | Valence State |
|---|---|---|---|
| S2 (Annealed) | 1.45 ± 0.01 | 0.19 ± 0.08 | **Sn$^{4+}$** |
| S1 (pristine) | 1.47 ± 0.02 | 0.74 ± 0.02 | **Sn$^{4+}$** |
| Standard Sn powder | 0.01 ± 0.01 | 0.48 ± 0.04 | **Sn$^{4+}$** |
|  | 2.59 ± 0.01 | 0.29 ± 0.01 | **Sn$^{2+}$** |

Mossbauer spectroscopy also probes the local environment around a nucleous and are very sensitive to local atomic structure affected by grain boundaries and the defects such as dislocations, antisites, vacancies and interstitials. The isomer shift (IS) obtained from the Mossbauer spectra is almost same for both the samples. With respect to SnO$_2$, the isomer shift (IS) values of about 2.5 mm s$^{-1}$ or more are considered for Sn$^{2+}$ ions and Sn$^{4+}$ ions are expected to show IS values lower than 2.0 mm s$^{-1}$ [29]. The observed IS values of both S1 and S2 samples indicate the presence of tetrahedral coordinated Sn$^{4+}$ matching the result of Benedetto et al., in similar compounds [30]. However, there is considerable change in the quadrupole splitting values for both the samples, suggesting the different degree of distortion at Sn sites for pristince and annealed samples. The absence of isomer shift for Sn$^{2+}$ ions substantiate the possibility of Sn at Cu sites, forming Sn$_{Cu}$ defect. The observed large quadrupole splitting in S1 is possibly associated with the tetrahydral distortion/ amorphization of Sn$^{4+}$ ions, randomly distributed in the host CZTS matrix. The high atomic fraction of Sn observed in the S1 by EDX measurement also facilitates the possibility of Sn to reside at the interstitial/antisites or disorderly arranged causing amorphization. The QS value decreased substantially for high temperature annealed S2 sample. The relative lowering in QS value suggests that annealing resulted in higher degree of crystallization and thus reducing the disorder for Sn$^{4+}$ in the CZTS system. This is also consistent with XRD measurements, showing one order lower dislocation densities for S2 sample. The absence of Sn in +II valance state in the both S1 and S2 samples is also governed due to the presence of excess sulfur in the system, which restrict reduction of Sn$^{4+}$ to Sn$^{2+}$ to neutralize charge states [31].

**Electrical characterization**

Temperature dependent electrical measruements are carried out from 100 K to 293 K temperature range using a liquid nitrogen cooled cryostate chamber. Measured two probe resistance is normalized and is shown in Fig 5 (a, b) for S1 and S2 samples, respectively. The measured resistance is showing reduction with increase in temperature, confirming the semiconducting behavior for these CZTS thin films. The measured lower resistivity for CZTS thin films for sample S2 is in agreement with the observed improved crystallinity of S2 sample, as discussed earlier. The improved crystallinity reduces the grain boundary barriers as compared to that of not annealed CZTS thin films (S1 sample).

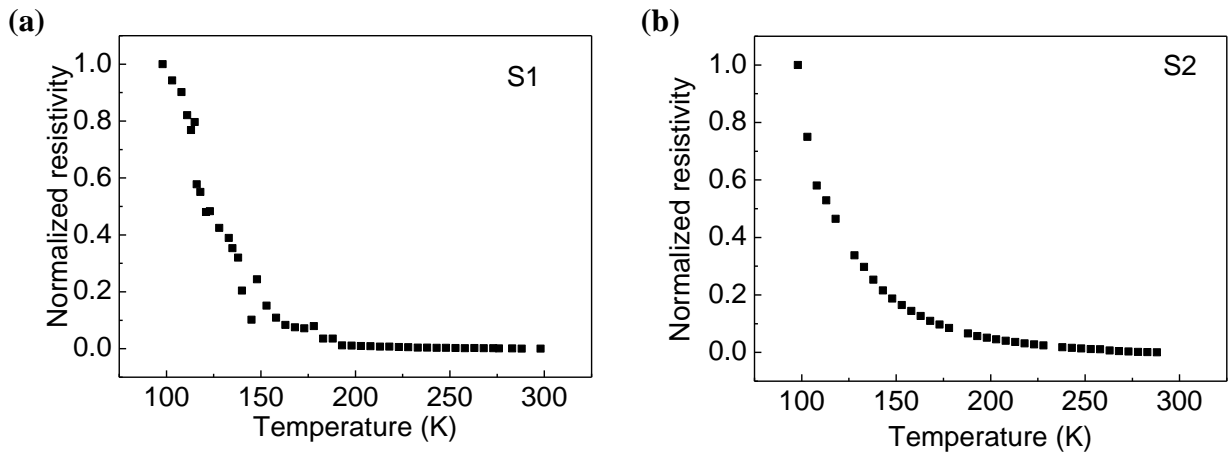

Figure 5: Normalized temperature dependent resistance variation with temperature for (a) S1 and (b) S2 CZTS samples

To understand closely the transport mechanism natural logarithm of normalized conductance is plotted with respect to inverse of temperature ( ln σ versus 1000/T ) from the resistivity vs. temperature data and is shown in Fig.6 (a, b) . We observed that ln σ versus 1000/T curves are comprised of two linear region for both S1 and S2 CZTS thin films. At high temperature region for T > 250 K the conduction in these films is governed by the thermionic emission and can be expressed as $\sigma = \sigma_0 exp(-E_a/KT)$, where $\sigma_0$ is a preexponential factor propotional to the grain size and average carrier concentration present in the film, $E_a$ is activation energy related to the barrier height in grain boundaries and '$K$' is the Boltzman constant [32].

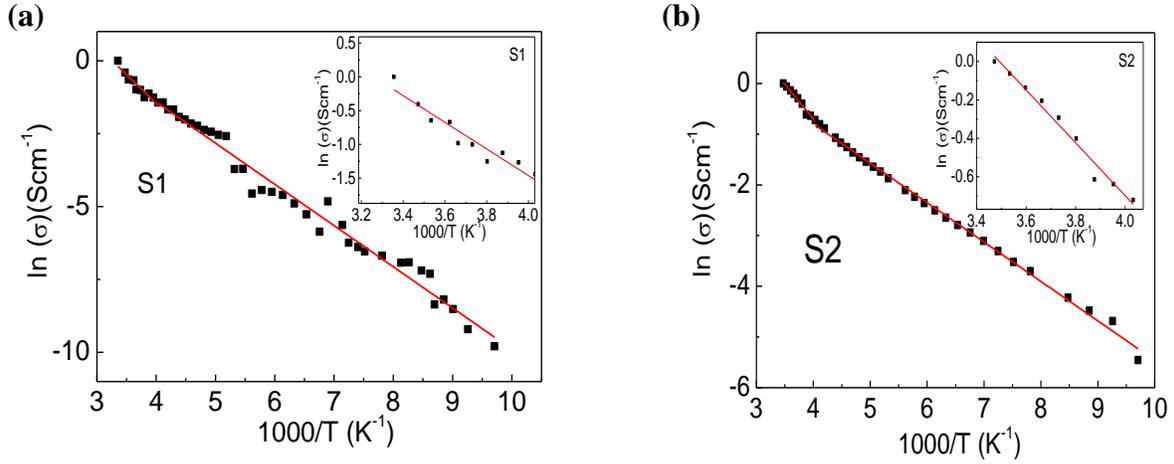

Figure 6: Natural logarithm of conductivity against 1000/T plots for (a) S1 and (b) S2 CZTS samples, with inset showing the high temperature regions for respective samples

Table 4. Activation energys for thermally activated and NNH carriers, as estimated from temperature dependent resistivity measurements.

| Sample | Slope1 | Slope2 | $E_a$ (meV) | $E_{NNH}$ (meV) |
| --- | --- | --- | --- | --- |
| S1 | 1.971 | 1.413 | 169.9 | 121.8 |
| S2 | 1.376 | 0.777 | 118.7 | 67 |

The activation energy of ~ 169.9 meV and 118.7 meV are measured from the repetive slopes in higher temperature region for S1 and S2 CZTS thin films, respectively. At lower temperature, holes in the p type semiconductor do not have sufficient energy to jump from acceptor level to the valance band and the conduction is dominated by the hopping of charge carriers. For low temperature of ln σ vs. 1000/T curve is fitted with a straight line considering nearest neighbour hopping (NNH) conduction mechanism. For NNH conduction in p type semiconductor hopping conductivity can be written as $\sigma = \sigma_{0NNH} exp(-E_{NNH}/KT)$, where $\sigma_{0NNH}$ is a preexponential factor, $E_{NNH}$ is activation energy for hopping conduction and $K$ is Boltzmann constant. The estimated activation energy for the NNH conduction are 121.8 meV and 67 meV for S1 and S2 CZTS thin films, respectively. These values are summarized in Table 4.

**Al:ZnO/i-ZnO/CdS/CZTS/Mo/SLG Device fabrication and characterization**

A complete solar cell device structure Al:ZnO/i-ZnO/CdS/CZTS/Mo/SLG is fabricated with low temperature processed (S1) and high temperature treated (S2) CZTS samples. The fabrication of complete device is reported elsewhere[27]. In brief, CZTS absorbers of thickness ~2.5 μm is deposited over Mo coated SLG substrate, followed by a thin layer of CdS of thickness ~50 nm using chemical bath deposition, which acts as the buffer layer and facilitates forming heterostructure p-n junction with the CZTS absorber layer. Thin intrinscic ZnO layer (~80 nm) is deposited using RF sputtering, as a window layer, which protects the bottom layer during the DC sputter deposition of aluminum doped ZnO layer working as the transparent top contact.

**Capacitance voltage characterization**

Capacitance-voltage (C-V) characteristics are measured under different bias potential at 5 kHz in dark for both pristine (S1) and annealed (S2) absorbers based photovoltaic devices. A relatively high frequency is considered intentionally to avoid the contribution of traps in the absorber layer in C-V measurements. The measured capacitance-voltage curves are shown in Fig 7 (a) for both the cells. The cell capacitance is dominated by the depletion or space charge capacitance at negative bias potential, whereas diffusion capacitance becomes significant with forward bias and an exponential increase in capacitance is observed with increasing voltage. The photovoltaic devices based on S1 and S2 absorbers exhibit different forward bias potentials for the exponential increase in capacitance values. This forward bias potential is relatively higher for S1 based absorber, suggesting that a higher potential is required for not annealed CZTS absorber based deivces. This is attributed to the relatively lower carrier concentration in pristine CZTS (S1 sample) as compared to that of post annealed CZTS (S2 sample) absorbers. Further, Mott-Schottkey plots (i.e plot of reciprocal of squaired capacitance at different bias potential) are shown in Fig 7 (b). The observed negative Mott-Schottky slopes substantiate p type conductivity for these CZTS films (both not annealed pristine (S1) and high temperature post annealed (S2) CZTS thin films). This Mott-Schottky slope is used to calculate the carrier concentrations and values are $6.95 \times 10^{16}$ cm$^{-3}$ and $2.51 \times 10^{17}$ cm$^{-3}$ for not annealed (S1) and annealed (S2) absorbers, respectively. Thus, observed relatively lower carrier concentration for S1 sample is in agreement with the observation that a large forward bias potential is required to drive the device into the diffusion capacitance region, Fig. 7(a).

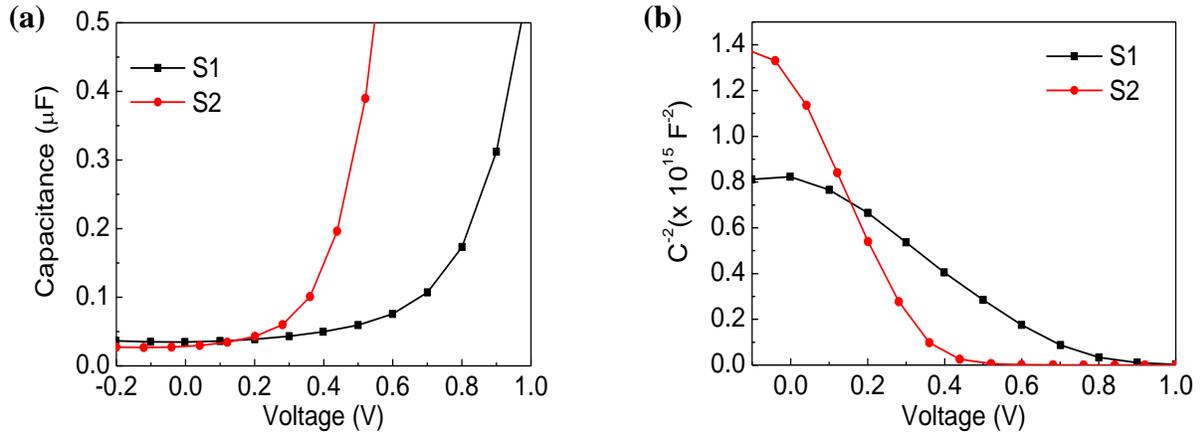

Figure 7: (a) Capacitance-Voltage (C-V) measurements (b) Mott-Schottky plots of photovoltaic devices based on S1 and S2 absorbers

The observed high carrier concentration for S1 as compared to S2 can be explained by the fact that tin based antisite defects acts as donor defects and serve to compensate the available most probable acceptor defects such as $Cu_{Zn}$ and $V_{Cu}$ present in the system. As there is no signature of tin in +II valance state is observed in Mossbauer measurements, the possibility of Sn at Cu sites is not probable. However, there is a high probability of Sn at Zn site because of lower formation energy of $Sn_{Zn}^{2+}$ in CZTS [33]. The increased conductivity of S2 is also in agreement with the theoratical reports suggesting that conductivity increases with increase in Cu/Sn ratio in CZTS samples[34].

**Photovoltaic characteristics**

The photovoltaic performance of solar cells is carried out under dark and one sun (AM1.5) illumination for both S1 and S2 absorbers and is shown in Fig 8 (a). The extracted device parameters are listed in Table 5. These observations suggest the improvement in efficiency for solar cell device fabricated using S2 absorber, which was subjected to the post annealing treatment. The obtained maximum efficiency with S1 and S2 are 0.145 % and 1.0%, respectively. The dark current densities are fitted with single diode model to obtain the reverse saturation current density ($J_0$) and the ideality factor (n) for these solar cell devices, Fig 8 (b). Ideality factor and reverse saturation current density of the solar cell are estimated using the diode equation $J = J_0 \left[ \exp\left(\frac{qV}{nKT}\right) - 1 \right]$, where $J$ is current density through the diode, V is voltage across the diode, $J_0$ is reverse saturation current density, q is electronic charge, $K$ is Boltzmann constant and T is absolute temperature. When cell is forward biased $J/J_0 \gg 1$, the current density can be expressed as $J = J_0 \exp\left(\frac{qV}{nK_BT}\right)$. The slope and the intercept of natural logaritham of current ($ln(J)$) versus voltage (*V*) curve provides ideality factor $n = \frac{q}{K_BT}\left(\frac{dV}{dln(J)}\right)$ and the reverse saturation current density can be extracted from $ln(J_0)$, respectively. The measured ($n$, $J_0$) values are (4.32, 23.85 µA/cm$^2$) and (2.17, 9.38 µA/cm$^2$) at lower forward bias voltage V< 0.4V for S1 and S2 solar cells, respectively. For higher bias voltage, the observed change in slope in the *lnJ vs V* characteristics is attributed to the enhanced carrier recombination because of increased minority carrier injection in the cell. Ideality factor greater than one substantiates the presence of unwanted recombination centers in the solar cell. We observed that S2 based solar cells exhibit lower recombination as compared to that of S1 absorber based solar cells.These results shows the detrimental impact of $Sn_{Zn}$ antisites defects, which are positioned deep in band gap and causes loss in open circuit voltage of the device resulting into inferior device performance[16]. Thus, these results supports the fact that Cu poor and Zn rich configuration of CZTS (S2) is relatively advantageous to realize enhanced solar photovoltaic response as compared to Cu poor and Sn rich configuration (in present case S1 sample).

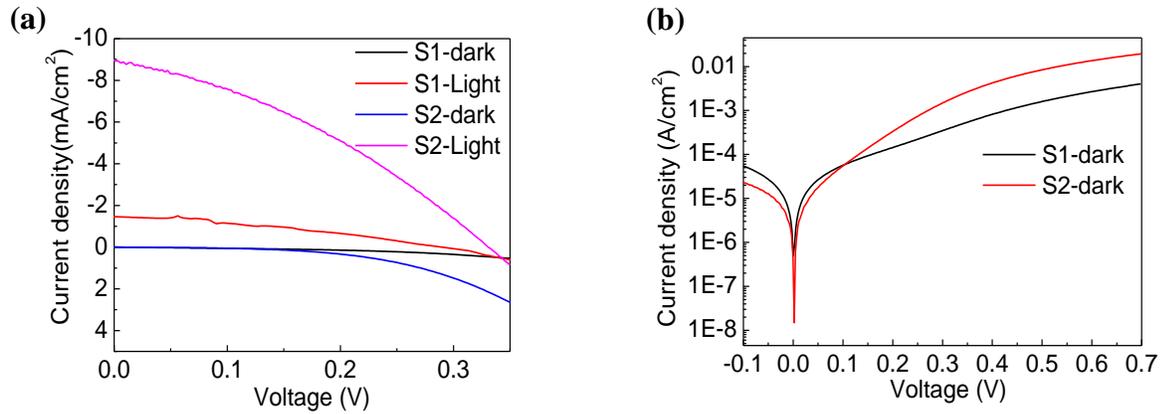

Figure 8: (a) Phorovoltaic characteristics of CZTS solar cell with S1 and S2 absorber in dark and under illumination (b) Dark IV characteristics of S1 and S2 solar cell

Table 5. Photovoltaic device parameters, estimated from the measured photovoltaic response for devices based on S1 and S2 abosrbers

| Sample | Voc (V) | Jsc (mA/cm$^2$) | FF (%) | Efficiency (%) | n | $J_0$ (μA/cm$^2$) |
|---|---|---|---|---|---|---|
| S1 | 285 | 1.46 | 35 | 0.145 | 4.32 | 23.85 |
| S2 | 331 | 9 | 35 | 1.0 | 2.17 | 9.38 |

**Conclusion**

A low cost sol-gel derived CZTS thin film synthesis process is used to achieve phase pure kesterite thin films. The low temperature processed sample S1 exhibited Sn rich stoichiometry, while sample S2, which was post annealing treatment at higher temperature, showed substantial Sn loss and changed into slightly Zn rich and Cu , Sn poor stoichiometry. Mossbauer measruements showed the presence of Sn at its preferred +IV oxidation states in both S1 and S2 and no signature of Sn$^{2+}$ state is noticed, thereby rulling out the possibility of $Sn_{Cu}$ donor defect. Large quadropple splitting observed in S1 is substantiate the large distortions at Sn site. Further, the excess Sn composition also facilitated the presence of Sn at Zn site, which is acting as a compensating donor defect to the available acceptor defects with low formation energy such as $Cu_{Zn}$ and $V_{Cu}$. The observed lower carrier concentration in S1 also validates this possibility. The post annealed S2 CZTS sample showed enhanced crystallization and reduced lattice distortion as evidenced from the smaller QS splitting in Mossbauer spectra. The lower concentration of Sn in S2 reduces the possibility of Sn to occupy Cu or Zn sites, which are beneficial for the photovoltaic performance. This is supported from the electrical and photovoltaic measurements, where solar cells based on S2 CZTS absorber showed enhanced photovoltaic response. The presented Mossbauer studies are very important to understand the available valance state of Sn in CZTS systems, which can be useful to properly investigate the defects and finally explaining the photovoltaic response.

**Acknowledgement:**

Author Ambesh Dixit acknowledges Department of Science and Technology (DST), Government of India, through Grant Nos. DST/INT/Mexico/P-02/2016 for carrying out the experimental work

**References:**

[1] P. Jackson, R. Wuerz, D. Hariskos, E. Lotter, W. Witte, M. Powalla, Effects of heavy alkali elements in Cu(In,Ga)Se2 solar cells with efficiencies up to 22.6%, Phys. Status Solidi - Rapid Res. Lett. 10 (2016) 583–586. doi:10.1002/pssr.201600199.

[2] B. Shin, Y. Zhu, N.A. Bojarczuk, S. Jay Chey, S. Guha, Control of an interfacial MoSe 2 layer in Cu 2ZnSnSe 4 thin film solar cells: 8.9 power conversion efficiency with a TiN diffusion barrier, Appl. Phys. Lett. 101 (2012). doi:10.1063/1.4740276.

[3] J. Tao, L. Chen, H. Cao, C. Zhang, J. Liu, Y. Zhang, L. Huang, J. Jiang, P. Yang, J. Chu, Co-electrodeposited Cu2ZnSnS4 thin-film solar cells with over 7% efficiency fabricated via fine-tuning of the Zn content in absorber layers, J. Mater. Chem. A. 4 (2016) 3798–3805. doi:10.1039/C5TA09636G.

[4] Y. Cao, M.S. Denny, J. V Caspar, W.E. Farneth, Q. Guo, A.S. Ionkin, L.K. Johnson, M. Lu, I. Malajovich, D. Radu, D.H. Rosenfeld, K.R. Choudhury, W. Wu, High-efficiency solution-processed Cu2ZnSn(S,Se)4 thin-film solar cells prepared from binary and ternary nanoparticles., J. Am. Chem. Soc. 134 (2012) 15644–7. doi:10.1021/ja3057985.

[5]     G. Larramona, S. Levcenko, S. Bourdais, A. Jacob, C. Choné, B. Delatouche, C. Moisan, J. Just, T. Unold, G. Dennler, Fine-Tuning the Sn Content in CZTSSe Thin Films to Achieve 10.8% Solar Cell Efficiency from Spray-Deposited Water-Ethanol-Based Colloidal Inks, Adv. Energy Mater. 5 (2015) 1–10. doi:10.1002/aenm.201501404.

[6]     T.K. Todorov, J. Tang, S. Bag, O. Gunawan, T. Gokmen, Y. Zhu, D.B. Mitzi, Beyond 11% Efficiency: Characteristics of State-of-the-Art $Cu_2ZnSn(S,Se)_4$ Solar Cells, Adv. Energy Mater. 3 (2013) 34–38. doi:10.1002/aenm.201200348.

[7]     S. Siebentritt, Why are kesterite solar cells not 20% efficient?, Thin Solid Films. 535 (2013) 1–4. doi:10.1016/j.tsf.2012.12.089.

[8]     H.S. Takuya Kato*, Homare Hiroi, Noriyuki Sakai, Satoshi Muraoka, CHARACTERIZATION OF FRONT AND BACK INTERFACES ON Cu2ZnSnS4 THIN-FILM SOLAR CELLS, Comp. A J. Comp. Educ. 27th Europ (2012) 1–3.

[9]     Y.S. Lee, T. Gershon, O. Gunawan, T.K. Todorov, T. Gokmen, Y. Virgus, S. Guha, Cu2ZnSnSe4 thin-film solar cells by thermal co-evaporation with 11.6% efficiency and improved minority carrier diffusion length, Adv. Energy Mater. 5 (2015). doi:10.1002/aenm.201401372.

[10]    W. Wang, M.T. Winkler, O. Gunawan, T. Gokmen, T.K. Todorov, Y. Zhu, D.B. Mitzi, Device characteristics of CZTSSe thin-film solar cells with 12.6% efficiency, Adv. Energy Mater. 4 (2014) 1–5. doi:10.1002/aenm.201301465.

[11]    J. Kim, H. Hiroi, T.K. Todorov, O. Gunawan, M. Kuwahara, T. Gokmen, D. Nair, M. Hopstaken, B. Shin, Y.S. Lee, W. Wang, H. Sugimoto, D.B. Mitzi, High efficiency Cu2ZnSn(S,Se)4 solar cells by applying a double in 2S3/CdS Emitter, Adv. Mater. 26 (2014) 7427–7431. doi:10.1002/adma.201402373.

[12]    C.J. Hages, N.J. Carter, R. Agrawal, T. Unold, Generalized current-voltage analysis and efficiency limitations in non-ideal solar cells: Case of Cu2ZnSn(SxSe1−x)4 and Cu2Zn(SnyGe1−y)(SxSe1−x)4, J. Appl. Phys. 115 (2014) 234504. doi:10.1063/1.4882119.

[13]    A. Crovetto, M. Palsgaard, T. Gunst, T. Markussen, K. Stokbro, M. Brandbyge, O. Hansen,

Interface band gap narrowing behind open circuit voltage losses in Cu$_2$ZnSnS$_4$ solar cells, 83903 (2017). doi:10.1063/1.4976830.

[14] P.D. Antunez, D.M. Bishop, Y. Luo, R. Haight, Efficient kesterite solar cells with high open-circuit voltage for applications in powering distributed devices, Nat. Energy. (2017) 1–7. doi:10.1038/s41560-017-0028-5.

[15] S. Kim, K.M. Kim, H. Tampo, H. Shibata, K. Matsubara, S. Niki, Improvement of a voltage deficit of Ge-incorporated kesterite solar cell with 12.3% conversion efficiency, Appl. Phys. Express. (2016) in press. doi:10.7567/APEX.9.102301.

[16] S. Chen, J.H. Yang, X.G. Gong, A. Walsh, S.H. Wei, Intrinsic point defects and complexes in the quaternary kesterite semiconductor Cu2 ZnSnS4, Phys. Rev. B - Condens. Matter Mater. Phys. 81 (2010) 35–37. doi:10.1103/PhysRevB.81.245204.

[17] S. Delbos, Kësterite thin films for photovoltaics : a review, EPJ Photovoltaics. 3 (2012) 35004. doi:10.1051/epjpv/2012008.

[18] T.J. Huang, X. Yin, G. Qi, H. Gong, CZTS-based materials and interfaces and their effects on the performance of thin film solar cells, Phys. Status Solidi - Rapid Res. Lett. 8 (2014) 735–762. doi:10.1002/pssr.201409219.

[19] D. Mutter, S.T. Dunham, Calculation of Defect Concentrations and Phase Stability in Cu2ZnSnS4 and Cu2ZnSnSe4 From Stoichiometry, IEEE J. Photovoltaics. 5 (2015) 1188–1196. doi:10.1109/JPHOTOV.2015.2430015.

[20] S. Bourdais, C. Choné, B. Delatouche, A. Jacob, G. Larramona, C. Moisan, A. Lafond, F. Donatini, G. Rey, S. Siebentritt, A. Walsh, G. Dennler, Is the Cu/Zn Disorder the Main Culprit for the Voltage Deficit in Kesterite Solar Cells?, Adv. Energy Mater. (2016) 1–21. doi:10.1002/aenm.201502276.

[21] S. Chen, X.G. Gong, A. Walsh, S.H. Wei, Defect physics of the kesterite thin-film solar cell absorber Cu 2 ZnSnS4, Appl. Phys. Lett. 96 (2010) 4–7. doi:10.1063/1.3275796.

[22] A. Polizzotti, I.L. Repins, R. Noufi, S.-H. Wei, D.B. Mitzi, The state and future prospects of

kesterite photovoltaics, Energy Environ. Sci. 6 (2013) 3171. doi:10.1039/c3ee41781f.

[23] S. Chen, L. Wang, A. Walsh, X.G. Gong, S. Wei, S. Chen, L. Wang, A. Walsh, X.G. Gong, S. Wei, Abundance of Cu Zn + Sn Zn and 2Cu Zn + Sn Zn defect clusters in kesterite solar cells Abundance of Cu Zn 1 Sn Zn and 2Cu Zn 1 Sn Zn defect clusters in kesterite solar cells, 223901 (2016) 10–14. doi:10.1063/1.4768215.

[24] H. Katagiri, K. Jimbo, M. Tahara, H. Araki, K. Oishi, The influence of the composition ratio on CZTS-based thin film solar cells Hironori Katagiri, Kazuo Jimbo, Masami Tahara, Hideaki Araki and Koichiro Oishi Nagaoka National College of Technology, 888 Nishikatakai, Nagaoka, Niigata 940-8532, Japan, Mater. Res. Soc. Symp. Proc. Vol. 1165. 1165 (2009) 1165-M04-1. doi:10.1016/j.solmat.2011.05.050.

[25] K. Biswas, S. Lany, A. Zunger, K. Biswas, S. Lany, A. Zunger, The electronic consequences of multivalent elements in inorganic solar absorbers : Multivalency of Sn in Cu 2 ZnSnS 4 The electronic consequences of multivalent elements in inorganic solar absorbers : Multivalency of Sn in Cu 2 ZnSnS 4, 201902 (2015) 94–97. doi:10.1063/1.3427433.

[26] G.K. Gupta, A. Dixit, Effect of precursor and composition on the physical properties of the low-cost solution processed Cu 2 ZnSnS 4 thin film for solar photovoltaic application, J. Renew. Sustain. Energy. 9 (2017) 13502. doi:10.1063/1.4974341.

[27] G.K. Gupta, A. Garg, A. Dixit, Electrical and impedance spectroscopy analysis of sol-gel derived spin coated Cu2ZnSnS4 solar cell, J. Appl. Phys. 13101 (2018) 0–11. https://doi.org/10.1063/1.5002619.

[28] G.K. Williamson, R.E. Smallman, III. Dislocation densities in some annealed and cold-worked metals from measurements on the X-ray debye-scherrer spectrum, Philos. Mag. 1 (1956) 34–46. doi:10.1080/14786435608238074.

[29] C.N. Banwell, Fundamentals of molecular spectroscopy, (1983) 338.

[30] F. Di Benedetto, G.P. Bernardini, D. Borrini, W. Lottermoser, G. Tippelt, G. Amthauer, 57Fe- and 119Sn- M??ssbauer study on stannite (Cu2FeSnS4)-kesterite (Cu2ZnSnS4) solid solution, Phys.

Chem. Miner. 31 (2005) 683–690. doi:10.1007/s00269-004-0430-y.

[31] S. Kim, J.S. Park, A. Walsh, Identification of Killer Defects in Kesterite Thin-Film Solar Cells, ACS Energy Lett. 3 (2018) 496–500. doi:10.1021/acsenergylett.7b01313.

[32] B.L. Guo, Y.H. Chen, X.J. Liu, W.C. Liu, a. D. Li, Optical and electrical properties study of sol-gel derived Cu2ZnSnS4 thin films for solar cells, AIP Adv. 4 (2014) 97115. doi:10.1063/1.4895520.

[33] S. Chen, A. Walsh, X.-G. Gong, S.-H. Wei, Classification of Lattice Defects in the Kesterite Cu(2) ZnSnS(4) and Cu(2) ZnSnSe(4) Earth-Abundant Solar Cell Absorbers., Adv. Mater. (2013) 1522–1539. doi:10.1002/adma.201203146.

[34] E. Garcia-Hemme, A. Fairbrother, L. Calvo-Barrio, E. Saucedo, I. Martil, Compositional Dependence of Chemical and Electrical Properties in $Cu_2ZnSnS_4$ Thin Films, IEEE J. Photovoltaics. 6 (2016) 990–996. doi:10.1109/JPHOTOV.2016.2566888.